\documentclass[utf8]{FrontiersinHarvard} 

\usepackage{url,hyperref,lineno,microtype,subcaption}
\usepackage[onehalfspacing]{setspace}
\usepackage{xcolor}
\usepackage{booktabs}
\usepackage{gensymb}
\usepackage{multirow}
\usepackage{comment}

\def\keyFont{\fontsize{8}{11}\helveticabold }
\def\firstAuthorLast{S. Youn {et~al.}} 
\def\Authors{SungWoo Youn\,$^{1,*}$, Junu Jeong\,$^{1}$ and Yannis K. Semertzidis\,$^{1,2}$}
\def\Address{$^{1}$Center for Axion and Precision Physics Research, Institute for Basic Science, Daejeon, Republic of Korea\\
$^{2}$Department of Physics, Korea Advanced Institute of Science and Technology, Daejeon, Republic of Korea}
\def\corrAuthor{SungWoo Youn}

\def\corrEmail{swyoun@ibs.re.kr}

\begin{document}
\onecolumn
\firstpage{1}

\title[Axion haloscope development at CAPP]{Development of axion haloscopes for high-mass search at CAPP} 

\author[\firstAuthorLast ]{\Authors} 
\address{} 
\correspondance{} 

\extraAuth{}

\maketitle

\begin{abstract}
The axion offers a well-motivated solution to two fundamental questions in modern physics: the strong CP problem and the dark matter mystery.
Cavity haloscopes, exploiting resonant enhancement of photon signals, provide the most sensitive searches for axion dark matter in the microwave region.
However, current experimental sensitivities are limited to the $\mathcal{O}(10^0)\,\mu$eV range, while recent theoretical predictions for the axion mass favor up to $\mathcal{O}(10^2)\,\mu$eV, suggesting the need of new experimental approaches that are suitable for higher mass regions.
CAPP has developed/proposed several haloscopes effective for high-mass axion searches based on new cavity concepts and practical tuning mechanisms.
They are characterized by large detection volumes and/or high quality factors at high frequencies, achieved by partitioning a single cavity into multiple cells, exploiting higher-order resonant modes, and constructing dielectric photonic crystal structures.
Improving on the dish antenna haloscope scheme, a horn antenna array has also been proposed for volume-efficient broadband search in the THz region.
We review these haloscope designs for sensitive search in the high-mass regions and discuss their impacts on future experiments.


\tiny
 \keyFont{ \section{Keywords:} Axion, dark matter, haloscope, cavity, high-mass} 
\end{abstract}

\section{Introduction}


Dark matter, one of the biggest mysteries of modern science, is believed to exist as a missing component of matter in our universe.
There are several candidates with very different identities depending on the mass, spanning an incredibly wide range~\citep{annurev-astro-082708-101659, Green_2021}.
Among them, the axion, originally proposed to solve the charge-parity symmetry problem in quantum chromodynamics (QCD) of particle physics~\citep{PecceiQuinn:1977,Weinberg:1978,Wilczek:1978}, has emerged as the preferred particle to explain the missing matter~\citep{Preskill:1983,Abbott:1983,Dine:1983}.
The invisible axions with very light masses are generally accepted as QCD axions and two main benchmarks, refereed to as the Kim–Shifman–Vainshtein–Zakharov (KSVZ)~\citep{Kim:1979,Shifman:1980} and Dine–Fischler–Srednicki–Zhitnitsky (DFSZ)~\citep{Zhitnitsky:1980,Dine:1981} models, are common targets of searches nowadays.
From a cosmological perspective, they are expected to behave like coherently oscillating waves due to their light mass and feeble interactions, making them suitable for forming dark matter halos.
Because of the large number density, when they are virialized, their velocity follows a Maxwellian distribution of $v\sim10^{-3}c$, resulting in a spectral bandwidth of $Q\sim10^{6}$~\citep{Turner:1990}.

Over the past few decades, there has been an intense experimental effort to explore the vast parameter space of axions and axion-like particles~\citep{cajohare:2020,Yannis:2022}.
One of the most popular detection principles relies on the electromagnetic (EM) interaction, the process by which an axion decays into two photons.
Experimental searches typically employ a strong magnetic field to provide a sea of virtual photons with which axions interact to be converted into real photons, an experimental signature~\citep{Sikivie:1983}.
Numerous axion haloscopes also involve microwave cavities to take advantage of the resonant enhancement of the signal when the photon frequency matches the cavity eigenfrequency~\citep{Sikivie:1985}.
The expected axion-photon conversion power from a cavity haloscope is given by
\begin{equation}
    P_{a\gamma\gamma} = 1.15 \times 10^{-23}\,{\rm W} \left(\frac{g_{\gamma}}{0.97}\right)^2 
    \left(\frac{\rho_a}{0.45\, {\rm \frac{GeV}{cm^3}}}\right) 
    \left(\frac{\nu_a}{5.3\,{\rm GHz}}\right) 
    \left(\frac{\langle \mathbf{B}_{e}^{2} \rangle}{\left(9.8\,{\rm T}\right)^2}\right)
    \left(\frac{V_{c}}{1.4\,{\rm L}}\right) 
    \left(\frac{Q_c}{10^4}\right)
    \left(\frac{G}{0.6}\right), 
    \label{eq:conv_power}
\end{equation}
where $g_{\gamma}$ is the coupling constant of the axion-photon interaction with values of $-$0.97 and 0.36 for the KSVZ and DFSZ models, respectively, $\rho_{a}$ and $\nu_a(=m_a/2\pi)$ are the local density and Compton frequency (mass) of the dark matter axion.
As the experimental parameters, $\langle \mathbf{B}_{e}^{2} \rangle$ is the average of the square of the external magnetic field inside the cavity volume $V_c$ and $Q_{c}$ is the cavity quality factor.
The form factor $G$, defined as
\begin{equation}
G = \frac{ |\int_{V} \mathbf{E}_c \cdot \mathbf{B}_e d^3x |^2} {\int_{V} |\mathbf{B}_e|^2 d^3x \int_{V} \epsilon|\mathbf{E}_c|^2 d^3x },
\label{eq:form_factor}
\end{equation}
with $\mathbf{E}_{c}$ being the electric field of the cavity resonant mode and $\epsilon$ the dielectric constant inside the cavity, is a mode-dependent parameter that represents the EM energy density induced by the axion electrodynamics.
For a cylindrical geometry, the TM$_{010}$ mode gives the maximum.

The cavity-based haloscope technique provides the most sensitive method for detecting axion signals, especially in the microwave region~\citep{DePanfilis:1987,Hagmann:1990,ADMX:2021,CAPP:2023,HAYSTAC:2023,Adair:2022}.
It is noted, however, that sensitive searches using cavity haloscopes are limited to a low frequency (mass) region, $\mathcal{O}(10^0)$\,GHz ($\mathcal{O}(10^0)\,\mu$eV).
This is mainly because the resonant frequency of the TM$_{010}$ mode of a cylindrical cavity is dictated by its radius $r$, i.e., $\nu_{{\rm TM}_{010}} = \frac{11.47\,{\rm GHz}}{r\,[{\rm cm}]}$, which results in a substantial loss of detection volume and a deterioration of cavity quality factor in high-frequency searches.
A linear increase of quantum noise with frequency also degrades the signal-to-noise ratio. 
For these reasons, only a few experimental attempts have been made to surpass this frequency region, and also the sensitivity is still far from that of the QCD axion model~\citep{Alesini:2022,Quiskamp:2022}.
Meanwhile, recent progress has been made in the theoretical aspect of predicting axion mass according to cosmological scenarios of axion production and evolution.
In particular, in the post-inflation scenario, the axion mass can be uniquely constrained by QCD lattice calculations and/or numerical simulations, favoring higher mass regions of tens to hundreds of $\mu$eV~\citep{Kawasaki:2015,Bonati:2016,Klaer:2017,Dine:2017,Buschmann:2022}.
Therefore, pushing the experimental sensitivity to this region is very important not only for axion search but also for testing the axion cosmology.
Over the past few years, the Center for Axion and Precision Physics Research (CAPP) at the Institute for Basic Science has proposed several haloscope designs that can effectively increase the search mass while significantly improving experimental performance compared to existing approaches.
The groundbreaking ideas include 1) multi-cell configuration of the resonator, 2) tuning mechanism for higher resonant modes, 3) implementation of photonic crystal structure, and 4) horn antenna array configuration.
In this article, we review these novel designs for sensitive searches for high-mass axions and discuss their implications for the next-generation experiments.

\section{High-performance strategies at high frequencies}

The main challenge of cavity haloscope search at high frequencies stems from the relationship between cavity performance and search frequency: the resonant frequency of the TM$_{010}$ mode is inversely proportional to the cavity radius, reducing the cavity volume and quality factor.
The conversion power decreases with the cavity resonant frequency $\omega_{c}(=2\pi\nu_c)$ by
\begin{equation}
\label{eq:scan_rate_for_omega_c}
    P_{a\gamma\gamma} \propto \omega_{c}^{-8/3},
\end{equation}
and thus, for example, doubling the search frequency reduces the conversion power by a factor of 6 for a fixed aspect ratio of the cavity.
Several ideas have been proposed to overcome this bottleneck in high-frequency search: 1) configuring multiple small resonators~\citep{Hagmann:1990b}, 2) exploiting higher resonant modes~\citep{McAllister:2018,Alesini:2021}, 3) implementing metamaterials with conductor arrays~\citep{Lawson:2019}, 4) even new haloscope techniques using reflective surfaces,~\citep{MADMAX:2017} and so on.

One of the most intuitive ways to make efficient use of a given magnet volume is to pack multiple small cavities together and combine their individual outputs while ensuring coherence of the axion signal~\citep{Hagmann:1990b}.
A subsequent study has shown that configuring a readout chain in which signal combination precedes amplification reduces the complexity of experimental design without considerable reduction of the signal-to-noise ratio~\citep{Jeong:2018}.
However, the resonant frequency of individual cavities must be matched with high precision, which becomes difficult as cavity multiplicity increases.
Moreover, this cavity configuration is still inefficient in using the magnet volume, primarily due to the unavoidable unused volume between cavities, related to packing efficiency.
An alternative design is to create multiple identical cells by dividing the internal volume of an existing cavity that fits into the magnet bore using equidistantly spaced metal partition~\citep{Jeong:2018b}.
Dubbed pizza cavity, this multi-cell design increases the resonant frequency while minimizing the volume loss.

Exploiting higher-order modes is also beneficial because it naturally increases the search frequency without volume loss and with higher quality factors~\citep{Alesini:2021} as seen in table~\ref{tab:higher-modes}. 
However, this approach has been discarded because the high-degree variations in the cavity EM field lead small form factors and difficult frequency tuning.
The problem can be resolved by strategically placing (periodic) structures of dielectric material to suppress the out-of-phase oscillating electric field components, which substantially improves the form factors~\citep{MADMAX:2017}.
For the TM$_{0{\rm n}0}$ modes of a cylindrical cavity, a layer of dielectric hollow can be considered for this purpose~\citep{McAllister:2018,Alesini:2021}.
Additionally, the property that the resonant frequency is strongly dependent on the thickness of the structure can be utilized to design an appropriate tuning mechanism~\citep{Kim:2020}.
\begin{table}[h]
\centering
\begin{tabular}{c|cccc}
\toprule
~~~Mode~~~ & $\nu_c$ & $~~~V_c~~~$ & $Q_c$ & $~~~G~~~$ \\
\hline
TM$_{010}$ & $\nu_{010}$ & $V_{010}$ & $Q_{010}$ & 0.69 \\ 
TM$_{020}$ & $2.3\times \nu_{010}$ & $V_{010}$ & $1.5\times Q_{010}$ & 0.13 \\ 
TM$_{030}$ & $3.6\times \nu_{010}$ & $V_{010}$ & $1.9\times Q_{010}$ & 0.05 \\ 
\bottomrule
\end{tabular}
\caption{Cavity parameters for different resonant modes of a cylindrical geometry. 
The values of $\nu_c$, $V_c$, and $Q_c$ are relative to the TM$_{010}$ mode, while those of $G$ are absolute.}
\label{tab:higher-modes}
\end{table}

Meanwhile, the periodic configuration of dielectric disks or metal wires has recently been considered as a new strategy for high-frequency search with large conversion volume.
The former utilizes the constructively boosted EM waves generated by the axion-induced field on the surface of high-$\epsilon$ dielectric disks precisely aligned in one dimension~\citep{MADMAX:2017}, while the latter exploits the coupling of axions to bulk plasmons in metamaterials composed of thin conducting wires arranged in two dimensions~\citep{Lawson:2019}.
The key advantage of the both schemes is that the search frequency is independent of the physical size of the detector, allowing large conversion volume and thus significantly increasing the signal produced at high frequencies.
However, the dielectric boosting system would require a new large-scale experiment to realize with reasonable sensitivity, and the plasma haloscope would pose some challenges in cavity construction and frequency tuning due to the large number of conducting thin wires that need to be handled.
A two-dimensional lattice structure of dielectric rods, forming a photonic crystal, can address those issues while retaining the benefit of large detection volume at high frequencies.
An auxetic behavior of rotating polygons, which enables isotropic expansion or contraction of the dielectric array in two dimensions, can be employed as a practical frequency tuning mechanism~\citep{Bae:2023}.

For exploration beyond tens of GHz, a dish antenna scheme is frequently considered based on the the concept that dark matter axions traversing a metallic surface under a magnetic field emit photons perpendicular to the surface~\citep{Jaeckel:2013,Horns:2013}.
While the methodology does not leverage resonant effects, it allows for broad range searches.
Recently, an interesting idea was proposed to strategically configure a parabolic reflector inside a conductive cylinder to focus the photons induced by axions on the lateral surface of the cylinder, offering the possibility of realizing dish antenna haloscopes in cylindrical geometries~\citep{Liu:2022}.
However, this approach makes limited use of the available volume within the magnetic field.
An array of horn antennas packed in a solenoid bore can greatly increase the surface area for a given magnet volume, significantly enhancing the conversion efficiency~\citep{Jeong:2023}.

\section{Haloscope development at CAPP}

\subsection{Multi-cell cavity}
\begin{figure}
\begin{center}
\includegraphics[width=0.55\linewidth]{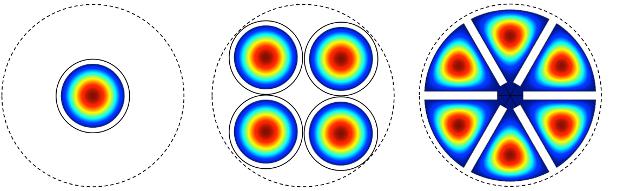}
\end{center}
\caption{Electric field distribution of the TM$_{010}$(-like) mode for a single cavity (left), a 4-cavity detector (middle), and a 6-cell cavity (right) within a given magnet bore (dotted circles). 
They are configured to have a similar resonant frequency.}
\label{fig:multi-cell}
\end{figure}

The multi-cell cavity is one of the haloscope concepts developed by CAPP as a volume-efficient approach for high-mass axion search~\citep{Jeong:2018b}.
It is characterized by multiple identical cells divided by equidistant thin metal partitions placed in a conventional cylindrical cavity.
This design is superior to the traditional multiple-cavity concept in terms of volume usage, as visually compared in figure~\ref{fig:multi-cell}.
In particular, the key idea is to introduce a partition gap in the center of the cavity. 
This allows all cells to be spatially connected to each other, enabling a single antenna to capture signals across the entire volume, simplifying the readout chain design.
Due to the symmetric structure, the pick-up antenna can be located at the cavity center, where only the TM$_{010}$ mode is detected when the system exhibits perfect symmetry.
Imperfect symmetry leads to asymmetric field distribution across the cavity volume, degrading the form factor.
This can be indicated by a non-vanishing electric field for the higher TM$_{0n0}$ (with $n>1$) modes, and the corresponding field strength estimates the form factor degradation.

Similar to the traditional mechanism for cylindrical cavities, frequency tuning can be achieved by introducing identical dielectric rods, one for each cell, and translating them along the azimuthal direction with respect to the cavity center.
Each tuning rod is extended out of the cavity through an arch-shaped opening made at the top and bottom of the cell.
The set of tuning rods is gripped at each end outside the cavity by a pair of multi-arm structures that rotate in a carousel configuration, allowing all the rods to move simultaneously.
The radius of the rod and the tuning path are optimized based on simulations using \texttt{COMSOL Multiphysics}~\citep{comsol} by maximizing the figure of merit defined as ${\rm F.O.M.}\equiv{\int }_{{\rm{\Delta }}\nu}({V}^{2}{C}^{2}Q)d\nu$ to obtain high cavity performance over a reasonable tuning range $\Delta\nu$.
An extended study has shown that the optimal rod size for multi-cell cavities matches that for cylindrical cavities of the same frequency and that the corresponding F.O.M. values are approximately the same regardless of the cell multiplicity~\citep{Jeong:2023c}. 

A potential issue of the multi-cell cavity design is related to field localization owing to geometric asymmetry, mainly arising from tolerances in cavity fabrication, including machining and assembly.
This gives rise to different field distributions in individual cells and eventually degrades the overall form factor.
The original cavity design consists of multiple identical bodies, each machined as a single piece with a central divider (partition) to form a multi-cell structure when assembled~\citep{Jeong:2018b}.
The monolithic structure makes precision machining difficult and thus the assembled cavity becomes vulnerable to forming a symmetric configuration.
We found that increasing the partition gap at the center of the cavity causes the individual cells to become more strongly coupled, effectively relieving field localization.
However, a large gap draws more field into the center of the cavity and eventually reduces the resonant frequency, losing the desired benefit for high-frequency searches.
To compromise between these two aspects, an extensive simulation study was performed to find the optimal size of the gap.
The general behavior is that higher cell multiplicity requires larger gap to maximize the F.O.M.
The simulation results were validated for several multi-cell (2-, 4-, and 6-cell) cavities of commercially available oxygen-free copper.
With the tuning structure taken into account, the effects can be further reduced by introducing a thinner (thicker) rod in the cell with the highest (lowest) field strength.
From an experimental perspective, the field localization effect can be estimated from the degree of imbalance in the distribution of electric (magnetic) field strength measured for the individual cells at relatively the same location using a mono-pole (loop) antenna.

\begin{figure}
    \centering
    \includegraphics[width=\linewidth]{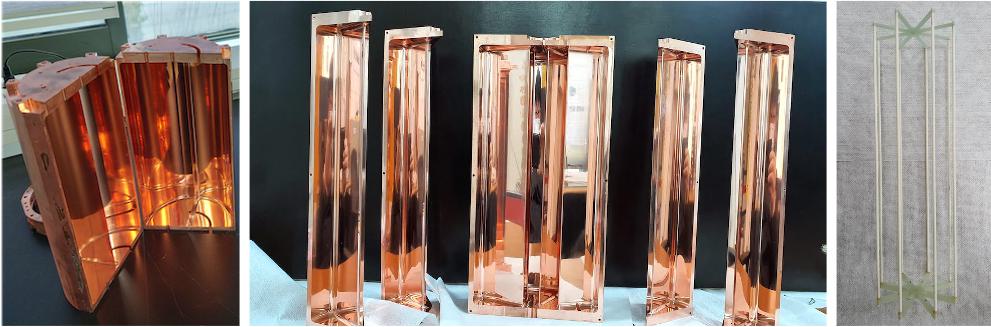}
    \caption{Photographs of individual pieces of the 2-cell and 8-cell cavities used for the experiments described in the text. 
    The tuning structure for the 8-cell cavity is also shown.}
    \label{fig:photo_multi-cell}
\end{figure}

The effectiveness of this unique design was successfully demonstrated by performing a series of axion search experiments at CAPP each with a multi-cell cavity implemented.
The first experiment was conducted using a double-cell cavity to prove the superiority of the multi-cell concept over the conventional one~\citep{Jeong:2020}.
The high performance and reliability enabled the search four times faster than a single cavity at the same frequency region and set new exclusion limits on the axion-photon coupling in the mass range between 13.0 and 13.9\,$\mu$eV.
Another experiment employed a larger cavity consisting of eight cells to search for higher-mass axion dark matter.
The high multiplicity increased the search frequency by a factor of 4 and the full use of a given magnet volume improved sensitivity, allowing the experiment to reach the KSVZ model at masses around 24.5\,$\mu$eV~\citep{Kutlu:2022So}.
Both experiments employed a set of alumina rods to tune the search frequency.
The resonant cavities used in these experiments are shown in figure~\ref{fig:photo_multi-cell}.
These results demonstrated that this new design could make considerable contributions to the next-generation searches for high-mass axions.

\begin{figure}[b]
\begin{center}
\includegraphics[width=0.45\linewidth]{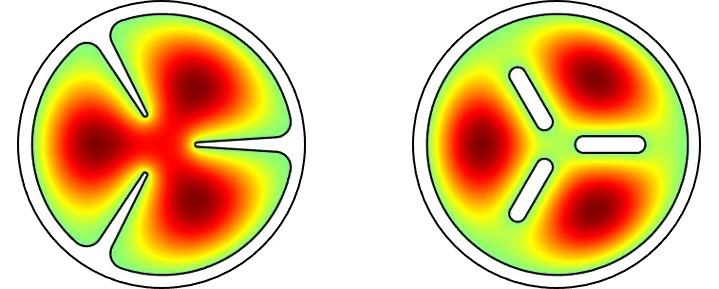}
\end{center}
\caption{Electric field distribution of the original (left) and modified (right) versions of a 3-cell cavity.}
\label{fig:3cell_kiwi}
\end{figure}

A modified version can be considered in which the partitions are detached from the monolithic structure and placed separately within the cavity, as shown in Fig.~\ref{fig:3cell_kiwi}.
This configuration reduces the complexity of cavity machining and thus improves the precision of cavity fabrication and assembly.
Each partition is slightly longer (about 100\,$\mu$m) than the cavity wall to create stronger contacts with the cavity end-caps.
Moreover, the additional space between the partition and cavity wall causes the individual cells to be more strongly coupled, making the field configuration more symmetrical, naturally reducing the field localization effect.
A simulation study has shown that despite a slight reduction in cavity quality factor, this design enhances the form factor and thus improves the overall performance of haloscope searches.
An experiment using a modified three-cell cavity is currently underway to search for axion dark matter in the mass range around 5.3\,GHz.

\subsection{Tuning mechanism for higher modes}
Higher-order resonant modes, as specified in table~\ref{tab:higher-modes}, are beneficial for increasing search frequency.
However, there are two major issues that need to be addressed: 1) the field structure contains counter-oscillating component(s) that reduces the form factor in a uniform external magnetic field and 2) there is difficulty in tuning the resonant frequency while maintaining the form factor at a reasonable level.
The former issue can be resolved by properly arranging a structure of dielectric material to suppress the counter-oscillating electric field component(s) and thereby restore the form factor~\citep{McAllister:2018,Alesini:2021}. 
For the TM$_{030}$ mode of a cylindrical cavity, for example, a layer of cylindrical dielectric hollow can be considered for this purpose.

\begin{figure}
    \centering
    \includegraphics[width=0.6\linewidth]{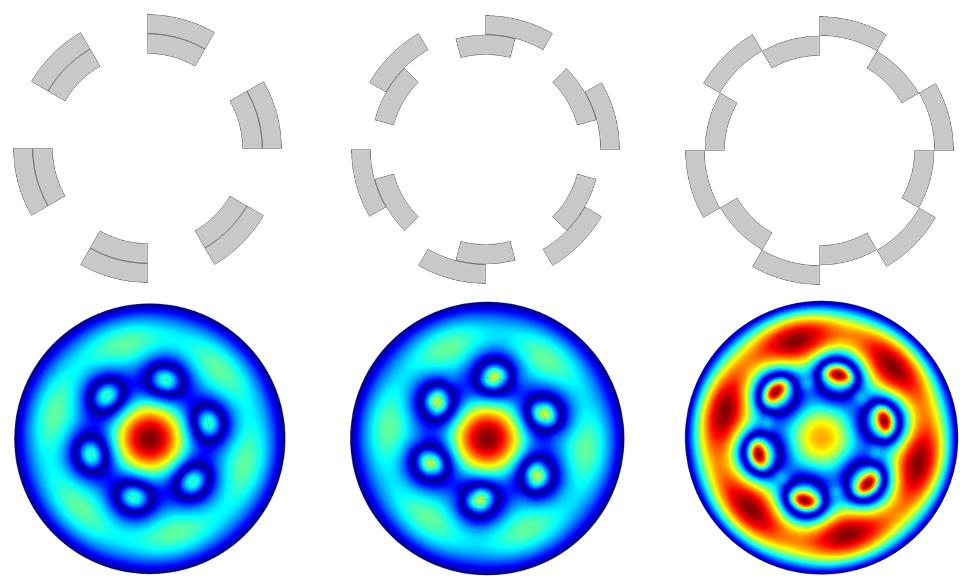}
    \caption{(Top) Illustration of the tuning mechanism using a two-layer segmented dielectric structure.
    The inner layer rotates relative to the outer layer, effectively varying the thickness of the structure.
    (Bottom) Corresponding electric field distributions.}
    \label{fig:wheel_mechanism}
\end{figure}

For frequency tuning, there are several schemes to be considered depending on the direction in which the field symmetry is broken (e.g., axial, radial, and azimuthal directions for a cylindrical geometry).
Breaking axial symmetry can be achieved by splitting the dielectric hollow into two parts horizontally and moving them along the axial direction. 
This was studied in Ref.~\citep{McAllister:2018}, especially for the TM$_{030}$ mode.
Similarly, the transverse symmetry can be broken by splitting the dielectric hollow into several pieces vertically and separating them along the radial direction. 
The scheme is described in Ref.~\citep{Alesini:2021}.
Symmetry breaking along the azimuthal direction relies on the EM characteristic in that the resonant frequency is strongly dependent on the thickness of the dielectric hollow.
Based on this property, we designed a new tuning mechanism consisting of a double layer of concentrically segmented dielectric hollow pieces in which one layer of the segments rotates with respect to the other~\citep{Kim:2020}, effectively varying the thickness of the structure, as visualized in figure~\ref{fig:wheel_mechanism}. 
For each scheme, the optimal dimensions of the tuning structure that maximize the F.O.M. were obtained based on extensive simulation studies. 
Among them, the one involving the symmetry breaking along the azimuthal direction showed the high sensitivity over a reasonable tuning range.
In this study, we did not consider the effects of non-minimal coupling between electromagnetism and gravity motivated by quantum phenomena such as vacuum polarization in extreme gravitational environments~\citep{Ravi_2023}.
While they could induce azimuthal dependence in axion-photon coupling~\citep{Chen2022, sakurai2023axion}, their impact was estimated to be negligible for laboratory axion haloscope experiments.

The third tuning scheme essentially relies on the transformation of continuous rotational symmetry into discrete rotation symmetry to create the tuning capability.
Since simultaneous rotation of the segments within a layer can be achieved using a single rotor, this mechanism is highly reliable and easy to implement.
The optimal number of segments and thickness/position of the layers for the ${\rm TM}_{030}$ mode were obtained through numerical simulations.
Using a model of a copper cylindrical cavity and dielectric segments of $\epsilon=10$, it was found that a pair of six-segment layers with thickness and inner radius of $\lambda/2\sqrt{\epsilon}$ ($\lambda$ is the wavelength of the EM wave) and $0.37R$ ($R$ is the cavity radius), respectively, gives the best performance.
We observed that the TM$_{030}$ resonant frequency can be tuned about 5\% and the form factor was estimated to be greater than 0.33 (compared to 0.05 for an empty cavity) over the entire tuning range, allowing reasonable sensitivity to be achievable.

\begin{figure}
    \centering
    \includegraphics[width=0.33\linewidth]{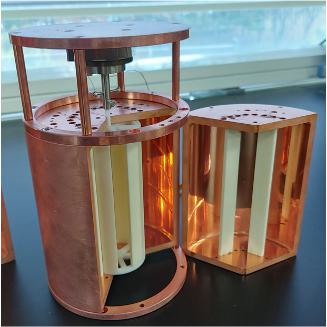}
    \includegraphics[width=0.45\linewidth]{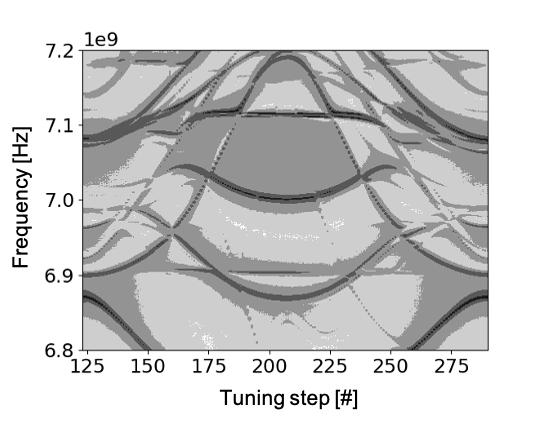}
    \caption{(Left) Photo of the partially assembled cavity with the tuning system mounted. 
    A piezo rotary actuator, positioned above the cavity, rotates the inner layer with respect to the outer layer, which remains fixed in position.
    (Right) Measured frequency map over a single tuning cycle. The bell-shaped curve with its tuning range between 6.9 and 7.2\,GHz corresponds to the TM$_{030}$ mode.}
    \label{fig:demo_wheel}
\end{figure}

The feasibility of this tuning mechanism was demonstrated using a copper cavity and a double-layer of alumina segments.
The cavity consisted of three identical pieces, each of which had two outer alumina (99.7\% Al$_2$O$_3$) segments placed at internal fixed positions.
Six smaller alumina segments supported by a pair of wheel-shaped structures at the top and bottom made up the inner layer of the tuning system.
The overall structure of the cavity and tuning system is shown in figure~\ref{fig:demo_wheel}. 
The cavity assembly was brought to a temperature of about 4\,K to test the tuning mechanism in a cryogenic environment.
By rotating the inner layer of the tuning system using a piezoelectric actuator, a mode map was drawn as a function of rotation angle.
We observed periodic patterns of the resonance modes during the continuous tuning process.
Figure~\ref{fig:demo_wheel} shows the mode map over a single cycle achieved with a rotation angle of 60$\degree$. 
The bell-shaped curve corresponds to our desired TM$_{030}$ mode with its frequency spanning from 6.9 to 7.2\,GHz, which is approximately three times higher than the TM$_{010}$ resonant frequency of the same cavity ($\nu_{010} = 2.5$\,GHz).
The smooth and symmetrical curve over $\sim 300$\,MHz indicates the stability and reliability of the tuning mechanism with a reasonable tuning range.
The measured quality factors ranged from 90,000 in the high-frequency region to 110,000 in the low-frequency region.
This confirmed that this unique design provides a practical search method for high-mass axion search that leverage the high-order resonant modes of cavity haloscopes.

\subsection{Photonic crystal cavity}

\begin{figure}
    \centering
    \includegraphics[width=0.5\linewidth]{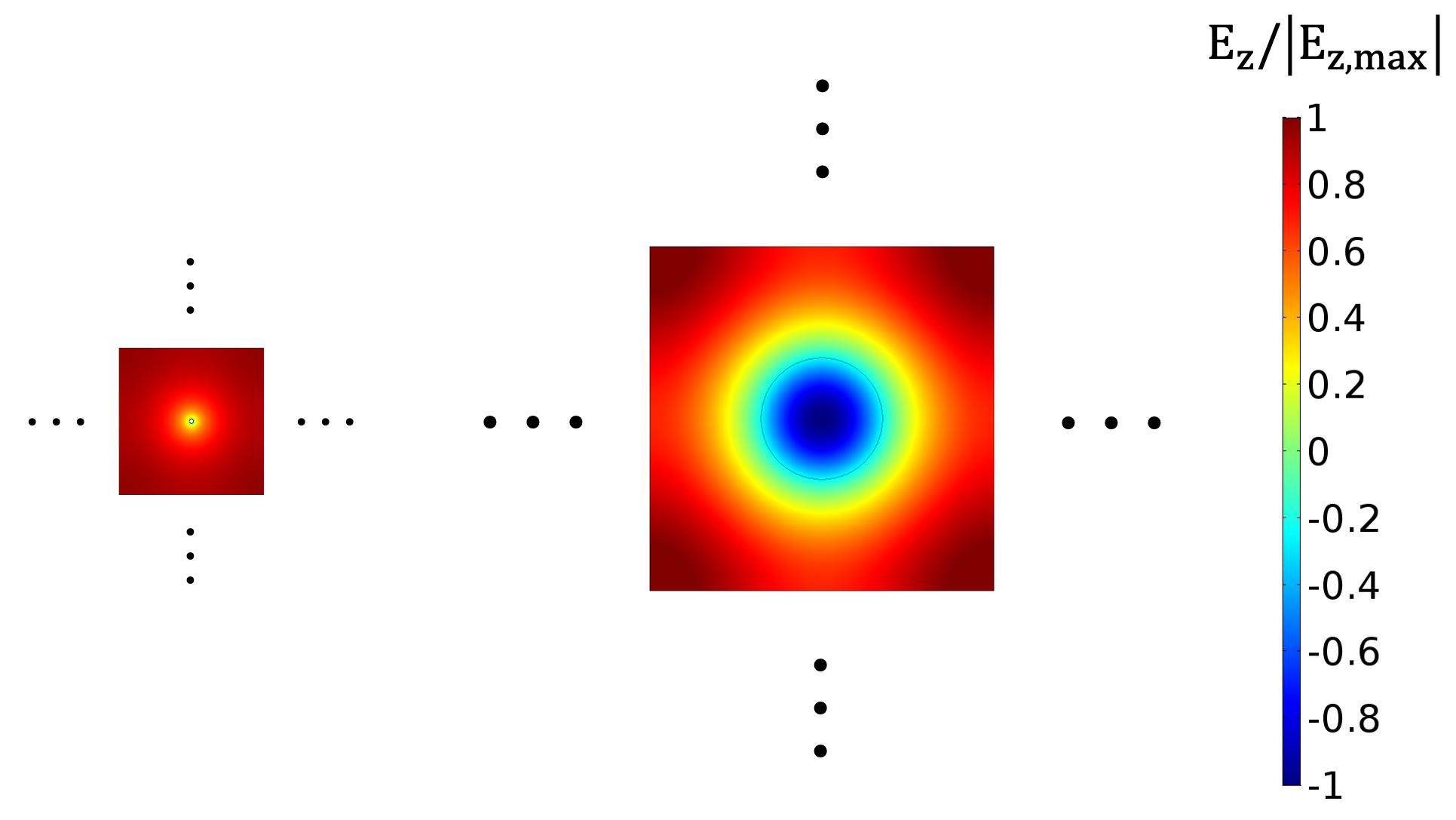}
    \caption{Normalized electric field distributions in a square unit cell with a conducting wire (left) and a dielectric rod (right) in the middle. 
    Two-dimensional lattice structures can be modeled as repeating patterns of the unit cells with periodic boundary conditions.}
    \label{fig:unit_cell}
\end{figure}

As discussed earlier, periodic (lattice) structures of dielectric or metallic materials can provide volume-efficient searches suitable for high-frequency regimes.
Motivated by the plasma haloscope scheme~\citep{Lawson:2019}, we explored a cavity design that incorporates an array of dielectric rods instead of thin conducting wires. 
This approach anticipates the use of low-loss dielectric materials to enhance the quality factor and make more efficient use of a given volume.
The two schemes, conducting wires vs. dielectric rods, were compared in terms of EM field properties using two-dimensional simulations.
A lattice structure was modeled by considering a square unit cell with the base material centered, as shown in figure~\ref{fig:unit_cell}, and arranging it in a two-dimensional repeating pattern with periodic boundary conditions imposed.
The dimensions of the unit cells and base materials were determined to have the same resonant frequency.
For the dielectric unit cell, the desired mode corresponds to the monopole mode of photonic crystals with a field component oscillating in opposite directions within the dielectric, resulting in a relatively low form factor.
However, the dielectric properties greatly improve the cavity quality factor in experimental environments, i.e., at low temperatures and high magnetic fields. 
Additionally, the larger size of the unit cell reduces the number density of the base material, facilitating cavity assembly and frequency tuning and reducing the mode population.

To evaluate the performance, a three-dimensional simulation study was performed with the lattice structures implemented in a common resonant cavity of $100\times100\times100$\,mm$^3$.
The wire metamaterial was simulated using a $10\times10$ array of 2\,mm-thick copper wires following the same structure as in Ref.~\citep{Millar:2023}, which gives the resonant frequency of 11.42\,GHz.
On the other hand, the photonic crystal was designed using a $6\times6$ array of 7.45mm-thick dielectric rods with $\epsilon=9.7$ to produce a similar resonant frequency.
The key quantity $V^2C^2Q$ was calculated numerically for various values of electrical conductivity ($\sigma$) and dielectric loss ($\tan\delta$), which are summarized in table~\ref {tab:performance_photonic}.
A 16-cell pizza cavity was also considered as reference.

\begin{table}[b]
    \begin{center}
    \begin{tabular}{cc|ccc}
    \toprule
    $\sigma$ [S/m] & $\tan\delta$     & Wire metamaterial & Photonic crystal & ~~~Multiple-cell~~~ \\
    \hline
    $6\times10^{7}$  & $1\times10^{-4}$ & $1.4\times10^{3}$ & $1.0\times10^{3}$ & $2.6\times10^{3}$ \\
    $6\times10^{8}$ & $1\times10^{-6}$ & $4.5\times10^{3}$ & $1.9\times10^{4}$ & $8.3\times10^{3}$ \\
    $~1\times10^{10}$ & $1\times10^{-8}$ & $1.8\times10^{4}$ & $9.0\times10^{4}$ & $3.4\times10^{4}$ \\
    \bottomrule
    \end{tabular}
    \caption{Numerically calculated $V^2C^2Q$ for different cavity designs in units of 10$^{-6}$\,m$^6$ for various values of $\sigma$ and $\tan\delta$.
    The first two rows represent commercially available copper and aluminum oxide at room and cryogenic temperatures, respectively, while the third row assumes superconductor and high-quality dielectric material.
    }
    \label{tab:performance_photonic}
    \end{center}
\end{table}

\begin{figure}
    \centering
    \includegraphics[width=0.6\linewidth]{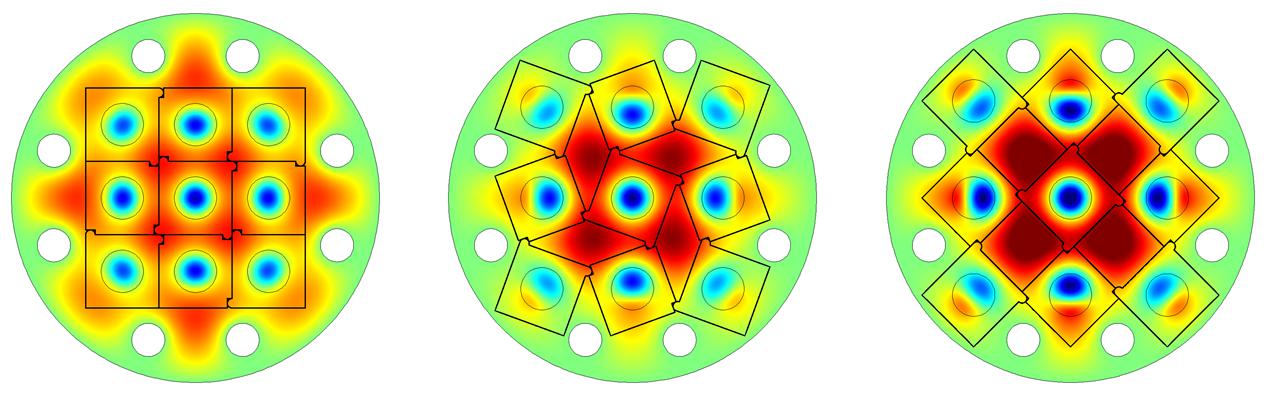}
    \caption{Two-dimensional design of a photonic crystal haloscope with a 3×3 square array (black square lines) in a conducting cylindrical cavity. 
    A dielectric rod is placed in the center of each square. 
    The array structure is deformed based on the auxetic behavior of the rotating rigid squares.
    The corresponding electric field distributions are displayed using the same color scale as figure~\ref{fig:unit_cell}. 
    Eight metal poles (white circles) are deployed to prevent field localization.}
    \label{fig:auxetic_tuning}
\end{figure}

Similar to the wire metamaterial haloscope, the resonant frequency of the photonic crystal haloscope can be tuned by varying the space between adjacent rods.
Therefore, an ideal frequency tuning mechanism requires expansion or contraction of the rod array at equidistant intervals in two dimensions.
We engaged the auxetic behavior of rotating rigid polygons to stretch or compress the array structure~\citep{Grima:2000}.
In particular, regular tessellation preserves a highly symmetrical pattern while rotating the unit cell to isotropically deform the structure.
For a realistic design, we modeled an array of 3x3 square blocks forming a regular tessellation with a dielectric rod anchored in the center of each block.
Each block has a different number of ears on the edges that connect it to adjacent blocks such that the overall structure expands when stretched and contracts when compressed, as illustrated in figure~\ref{fig:auxetic_tuning}. 
It is noteworthy that 1) the structure can be expanded or contracted by rotating the central block, allowing frequency tuning with a single rotating device, and 2) all the tuning rods move linearly in the radial direction with respect to the center, offering a straightforward way to guide them.
The size and spacing of the rods are determined to match the desired frequency, while the number of layers is chosen to accommodate the given cavity volume.

\begin{figure}
    \centering
    \includegraphics[width=0.25\linewidth]{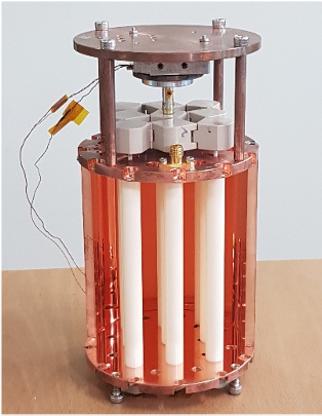}
    \includegraphics[width=0.5\linewidth]{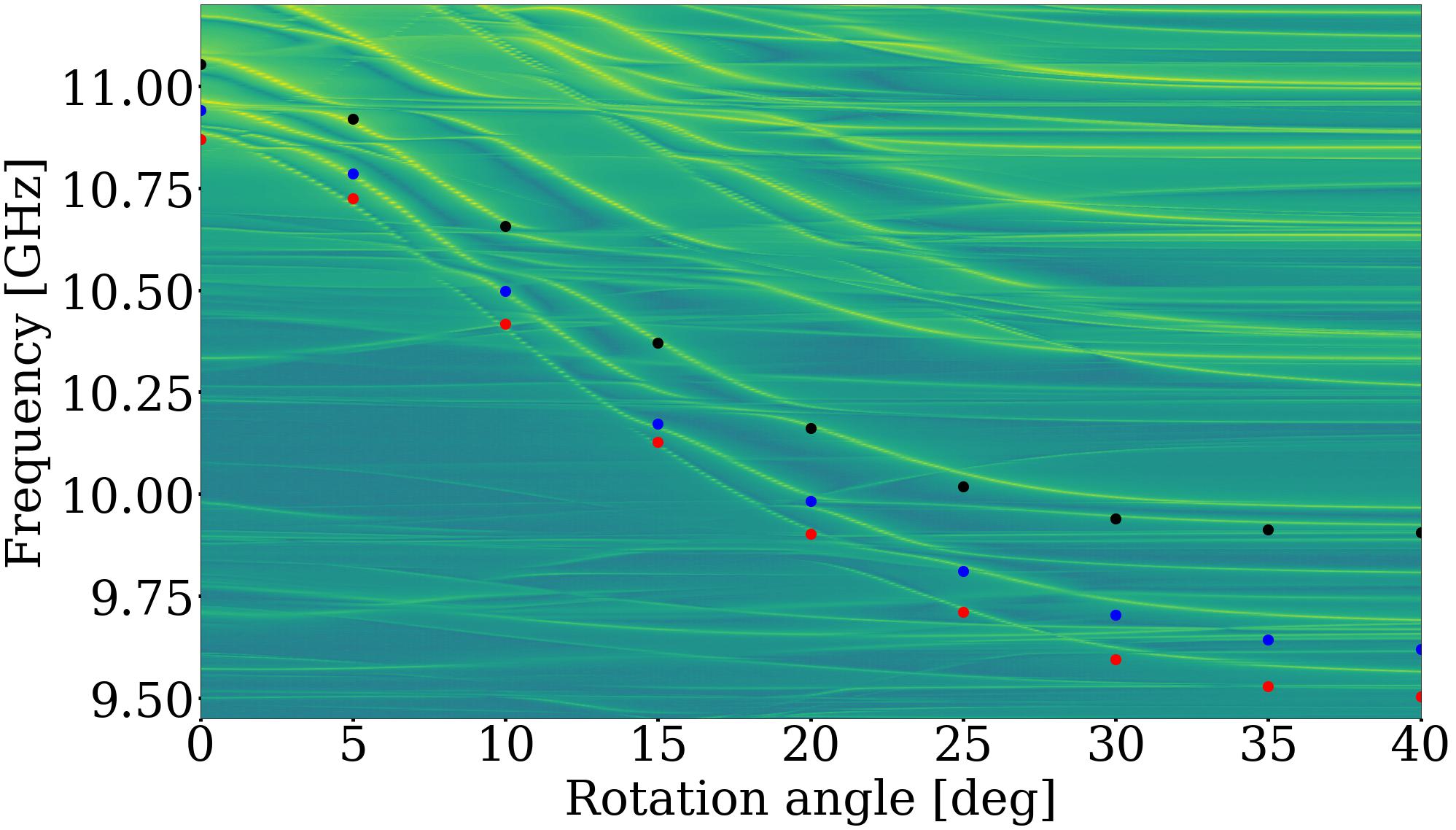}
    \caption{(Left) Photo of the  demonstration photonic crystal cavity depicted in the text.
    (Right) Mode map obtained at 4.5\,K, overlaid with the simulation indicated by the colored dots.
    Our desired mode corresponds to the lowest curve (with red dots overlaid).
    }
    \label{fig:demo_photonic}
\end{figure}

The concept of the photonic crystal haloscope was demonstrated using a cylindrical copper cavity and a set of alumina tuning rods.
The cavity had a conventional cylindrical shape with internal dimensions of $\O 78\,{\rm mm} \times 100\,{\rm mm}$.
The frequency tuning structure consisted of a $3\times3$ array of PEEK square blocks with a rod fixed at the center of each block.
A piezoelectric rotator was attached to the center block to induce auxetic motion of the array structure.
Since the tuning rods move only radially, as mentioned earlier, the end caps have slotted openings along the radial direction to guide their movement.
The overall structure of this demo cavity design can be seen in figure~\ref{fig:demo_photonic}.
To test the tuning mechanism in an experimental environment, the cavity assembly was installed in a vacuum chamber immersed in a reservoir of liquid helium.
Helium exchange gas was used to cool down the assembly to 4.5\,K.
A mode map was drawn by rotating the central block of the tuning structure through a full rotation angle (40$\degree$), as shown in figure~\ref{fig:demo_photonic}.
The desired mode corresponding to the lowest curve shows a frequency range of 10.9--9.6\,GHz, which is in good agreement with the simulations.
The cavity quality factor was measured during the tuning process, reaching the highest value of approximately 210,000 near 10.1\,GHz and slowly declining in both low and high frequency regions.
Additionally, we observed a 5\% increase in quality factor with a magnetic fields $\gtrsim 1$\,T, which is likely due to the similar dielectric property reported in Ref.~\citep{DiVora:2022}.
This demonstrated the experimental feasibility of the photonic crystal cavity design along with the newly invented tuning mechanism.

\subsection{Horn antenna array}

The cavity-based axion haloscopes, including those described in this text, would face considerable challenges when extending the search frequency well beyond several tens of GHz.
Entering this region demands a significant increase in multiplicity, which in turn leads to complexity and subtlety in building and operating systems, making them vulnerable to mode mixing and field localization.
A new technique was proposed to search for weakly interacting light particles with dish antennas relying on the detection principle that a reflective metal surface in a strong magnetic field effectively converts these particles into EM radiation emitted perpendicular to the surface~\citep{Jaeckel:2013,Horns:2013}.
The scheme is broadband and allows to explore a wide range of masses in a single measurement.
The axion-photon conversion power per unit area is given by
\begin{equation}
\label{eq:DAH}
    \frac{P_{a\gamma\gamma}}{A} \approx 1.3\times 10^{-26}\,{\rm W/m^{2}} \left( \frac{g_{\gamma}}{0.97}\right)^{2} \left( \frac{\rho_{a}}{0.45 \frac{\rm GeV}{\rm cm^{3}}}\right) \left( \frac{ B_{\parallel} }{10\,{\rm T}}\right)^{2},
\end{equation}
where $A$ is the area of the metal surface and $B_{\parallel}$ is the parallel component of the magnetic field to the metal surface.
Focusing these radiated photons enhances the experimental sensitivity in searching for dark matter axions over a broad frequency range~\citep{Bajjali:2023}.
Another research group proposed an interesting design that involves an outwardly focused parabolic mirror inserted in the center of a cylindrical conductor~\citep{Liu:2022}.
This unique geometry allows the dish antenna principle to be realized in standard cryostats and high-field solenoids and enables sensitive searches in the THz region that is difficult to access by cavity haloscopes.

\begin{figure}
    \centering
    \includegraphics[width=0.65\linewidth]{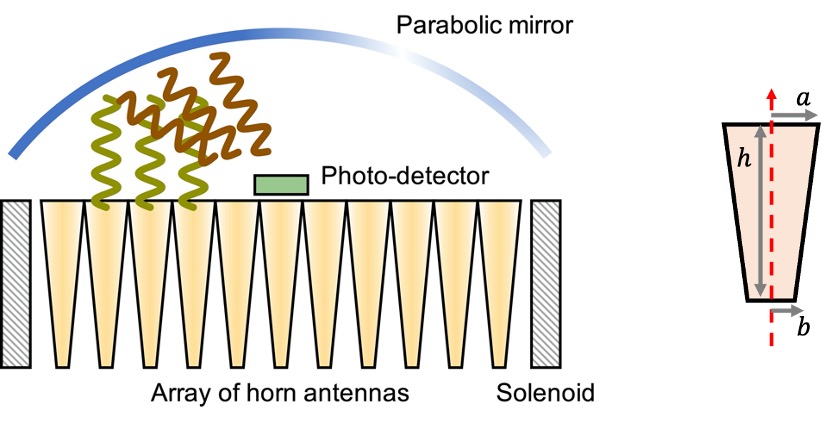}
    \caption{(Left) Schematic diagram of the horn-array haloscope. 
    The solenoid magnet (hatched black rectangles) encompasses the array of horn antennas (orange cones), which emit axion-induced photons (olive curly lines). 
    These photons are reflected (brown curly lines) by a parabolic mirror (blue curve) and focused on a photo-detector (green rectangle).
    (Right) The dimensions of an individual conical horn antenna are denoted by $a$, $b$, and $h$.}
    \label{fig:horn-array}
\end{figure}

One advantage of the dish antenna haloscope is that the radiation power is proportional to the surface area, making it easy to scale up.
Building upon this concept, CAPP has introduced an advanced geometric design for the dish antenna haloscope, aiming for volume-efficient searches for a given experimental setup~\citep{Jeong:2023}.
This design aimed to insert an ``array" of horn antennas, effectively increasing the surface area within the solenoid bore.
When low-aspect-ratio conical horn antennas are aligned parallel to the solenoid field, each antenna serves to convert dark matter axions into photons, thereby enhancing the total radiation power.
The size of the horn dictates both the minimum frequency and surface area of axion-photon conversion; a smaller aperture shifts the search frequency towards higher regions while accommodating more antennas.
The converted photons continue to reflect off the inner surface of each antenna until they eventually escape.
By positioning a dielectric lens at the antenna opening, the escaped photons can be deflected and aligned along the antenna axis.
These highly directional photons can be reflected by a parabolic mirror placed outside the solenoid magnet and then focused at a single point where a photo sensor is located.
A schematic of this proposed design is shown in figure~\ref{fig:horn-array}.

As the conversion power is directly related to the surface area within a magnetic field, the total power increases proportionally to the number of antennas inside the magnet volume.
For an array of $N$ identical conical horn antennas packed in a solenoid with bore radius $R$, the total radiation power is calculated as
\begin{equation}
\label{eq:Ptotal_hah}
\begin{split}
    P_{\rm tot} = N \times P_{\rm sgl} 
    \approx \Gamma \left( \frac{R}{a} \right)^{2} P_{\rm sgl}
    = 3.9\times 10^{-24}\,{\rm W} \left(\frac{g_{\gamma}}{0.97} \right)^{2} \left( \frac{B_{0}}{10\,{\rm T}}\right)^{2} \left(\frac{\Gamma}{0.75} \right) \left(\frac{R}{0.75\,{\rm m}} \right)^{2} \left(\frac{h}{2.1\,{\rm m}} \right),
\end{split}
\end{equation}
where $P_{\rm sgl}$ is the conversion power by a single antenna and $\Gamma$ is the packing efficiency.
We assume antennas dimensions with $a=10$\,mm and $b=1$\,mm, and allow a space of 1\,mm between adjacent antennas.
This conversion power level is comparable with that of typical cavity haloscopes.
We conducted a simulation study of axion electrodynamics using \texttt{COMSOL Multiphysics} to verify that the analytical approximation given by equation~\ref{eq:Ptotal_hah} is consistent with the numerical results, even at sufficiently high frequencies~\citep{Jeong:2023JKPS}.
Furthermore, we found that a simple hemispherical dielectric lens is adequate for aligning the emitted photons in the desired direction. 
Determining the optimal focusing geometry is left as a topic for future study.
The study also showed that a horn antenna with an aperture radius of 10\,mm is effective for experimental searches above 50\,GHz, and the proposed array design increases sensitivity to the axion-photon coupling by a factor of 30 compared to a single antenna configuration for the same magnet.

\section{Conclusion}
While cavity-based haloscopes represent the most sensitive method in searching for axion dark matter, their experimental sensitivity has been limited to the low $\mu$eV region.
In line with recent theoretical predictions favoring axion masses up to $\mathcal{O}(10^2)\,\mu$eV, it is crucial to extend the sensitive search to higher mass regions.
We reviewed the innovative haloscope designs developed by CAPP, which offer effective strategies for expanding the search frequency.
They include: 1) a pizza cavity design with multiple cells divided by equidistant partitions, tuned by a single carousel-like structure and readout by a single antenna; 2) a wheel tuning mechanism suitable for exploiting higher-order resonant modes such as TM$_{030}$; 3) a tunable photonic crystal design featuring an array of dielectric rods whose position can be adjusted using an auxetic structure of rotating squares; and 4) an array of horn antennas serving as a volume-efficient broadband haloscope.
The successful demonstrations indicate that these designs will provide significant advantages in the quest for high-mass axions with superior performance.
We stress that these haloscopes will pave the way for the next generation of experimental searches for invisible QCD axions, serving as an important test of axion cosmology.

\section*{Author Contributions}
SY: Writing - original draft, Conceptualization, Methodology, Supervision, Project administration;
JJ: Writing - editing \& review, Conceptualization, Software, Methodology;
YkS: Writing - review, Supervision, Project administration, Funding acquisition.

\section*{Funding}
This work was supported by the Institute for Basic Science (IBS-R017-D1-2023-a00).

\section*{Acknowledgments}
This is a short text to acknowledge the contributions of specific colleagues, institutions, or agencies that aided the efforts of the authors.

\bibliographystyle{Frontiers-Harvard}
\bibliography{frontiers}

\end{document}